\documentclass[a4,11pt]{article}
\usepackage{amssymb}
\usepackage{amsmath}
\usepackage[pdftex]{graphicx}
\usepackage{cite}
\usepackage{slashed}
\usepackage{bm}
\usepackage[top=2.5cm, bottom=2.0cm, left=2.5cm, right=2cm]{geometry}

\begin{document}

\begin{center}
{\large {\bf
Investigation of the $l^{-}l^{+}\nu \overline{\nu}$ final state at multi-TeV muon colliders through the exclusive decay of ZZ/WW gauge bosons in the Randall-Sundrum model}}\\

\vspace*{1cm}

{ Bui Thi Ha Giang$^{a, }$ \footnote{e-mail: giangbth@hnue.edu.vn, ORCID: 0000-0001-5814-0645}, Dang Van Soa$^{b}$}\\

\vspace*{0.5cm}
 $^a$ Hanoi National University of Education, 136 Xuan Thuy, Hanoi, Vietnam\\
 $^b$ Faculty of Applied Sciences, University of Economics - Technology for Industries,\\
 456 Minh Khai, Hai Ba Trung, Hanoi, Vietnam
\end{center}

\begin{abstract}
	
\hspace*{1cm} An attempt is made to present the effect of  the exclusive decay of ZZ/WW gauge bosons at high energy colliders in the Randall-Sundrum (RS) model. In this paper, we investigate the $l^{-}l^{+}\nu \overline{\nu}$ final state at muon-TeV colliders through the exclusive decay of ZZ/WW gauge bosons in detail. The result shows that with fixed collision energies, cross-sections for $l^{-}l^{+}\nu \overline{\nu}$ production in final state depend strongly on the parameters of the unparticle physics, muon polarization coefficients, parameters on anomalous couplings and also KK-graviton propagators. With the benchmark background $(\Lambda_{U}, d_{U})$ $= (1 $TeV$, 1.9)$, the total cross-sections achieve the maximum value when both of muon beams polarize left or right. In case of the different polarization, the cross section increases as the collision energy increases. The numerical evaluation shows that the cross-section for $l^{-}l^{+}\nu \overline{\nu}$ final state through the exclusive decay of WW charged bosons is much larger than that of ZZ neutral bosons under the same conditions. With the contribution of new physics in the RS model, the effect is greatly enhanced and the cross-sections for the production of $l^{-}l^{+}\nu \overline{\nu}$ final state can be measured in the future muon collisions.
\end{abstract}

\section{Introduction}
\hspace*{1cm} The Standard Model (SM) is the theoretical framework that describes successfully the experimentally observed phenomena of the fundamental particles and their interactions. Despite its many successes, the theory is imperfect and has some shortcomings. 
 The Randall-Sundrum (RS) model is one of the extended models based on the SM and solve the hierarchy problem naturally by the existence of warp factor in the metric \cite{rs, frank, dominici}. The RS model involves two 3-branes called ultraviolet (UV) and infrared (IR) branes. Throughout the popular RS1 scenario \cite{rs}, all the SM and Dark matter (DM) fields are considered to localized on the IR brane, whereas any other physics that is Plank-suppressed are on the UV brane \cite{folga}. There exists an additional scalar called the radion ($\phi$ ), corresponding to the quantum fluctuations of the distance between the two 3-branes. Due to the same quantum numbers, a possibility of mixing between the radion and the Higgs boson is considered \cite{frank, dominici, folga, ahm}. All phenomenological signatures of the RS model including the Higgs-radion mixing are determined by the parameters: the mixing parameter $\xi$, the vacuum expectation value (VEV) of the radion field $\Lambda_{\phi}$, the Higgs mass $m_{h}$ and the radion mass $m_{\phi}$. A complementary way to probe the mixing is the direct search for the new couplings exclusively allowed with a non-zero mixing parameter $\xi$ \cite{che}. In particular, there also exist Kaluza-Klein (KK) towers of massive spin-2 gravitons which interact with the SM fields\cite{mgfol, folga1, xiao}.\\
\hspace*{1cm} After the discovery of the Higgs at the Large Hadron Collider (LHC) at CERN \cite{geo,ser}, the particle content of the SM has finally been completed. In the Lagrangian of SM, the scale invariance is broken at or above the electroweak scale \cite{zhang, cheung}. At TeV scale, the scale invariant sector has been considered as an effective theory and that if it exists, it is made of unparticle suggested by Geogri \cite{georgi,georgi2} and may become part of reality. Based on the Banks-Zaks theory \cite{banks,chenhe}, unparticle stuff with nontrivial scaling dimension is considered to exist in our world and this opens a window to test the effects of the possible scalar invariant sector, experimentally. In the effective theory, below the BZ scale $\Lambda_{U}$, the form of the BZ operators matches unparticle operators and non-renormalizable operators as follows
\begin{equation}
\frac{C_{U}\Lambda_{U}^{d_{BZ} - d_{U}}}{M_{U}^{k}}  \mathcal{O}_{SM} \mathcal{O}_{U} = \dfrac{\lambda}{\Lambda_{U}^{d_{U}}} \mathcal{O}_{SM} \mathcal{O}_{U}.
\end{equation}
Here $d_{BZ}$ and $d_{U}$ are the scaling dimensions of the $\mathcal{O}_{BZ}$ and the unparticle $\mathcal{O}_{U}$ operators, respectively. The coefficient function $C_{U}$ is fixed by the matching condition. The parameter $\lambda = C_{U}\Lambda_{U}^{d_{BZ}}/M_{U}^{k}$ is the measure of the coupling between SM particles and unparticle. The range $1 < d_{U} < 2$ is the most natural since it is close to the particle limit of $d_{U} = 1$ \cite{ninh}. Moreover, unparticle effects are the largest in this region. Unparticle signals and new physics are expected to be seen at LHC when the bounds for $\Lambda_{U}$ are around 1 TeV \cite{jolee, alie}. The unparticle contributions to $B\rightarrow D^{*}\tau\nu$ have been investigated, the scalar unparticles are considered because contributions from vector unparticles are expected to be very small \cite{jolee}. From the current non-relativistic analysis, an essential conclusion can be drawn that unparticle effects might be tested in atomic physics experiments \cite{wondrak}. The effects of unparticle on properties of high energy colliders have been intensively studied in Refs.\cite{pra,alan,maj, kuma,sahi,kiku,chen,kha}. It is natural to ask whether an alternative perspective on unparticle can be formulated that is more transparent and more tractable a Yang-Mills hidden sector at strong coupling. A compelling framework to explore in this context is that of models incorporating warped extra dimensions. Therefore, the research the unparticle in RS framework, one of warped extra dimension model, is inevitable, e.g. Refs.\cite{fried, iltan, giang, soa1}. Search for DM and unparticles produced in association with a Z boson in proton-proton collisions at $\sqrt{s} = 8$ TeV has been presented in Ref.\cite{khac}. The anomalous couplings at LHeC is researched in Ref.\cite{giang2023}. The muon colliders, which can reach center-of-mass energy up to tens of TeV, provide an unprecedented potential in probing new physics beyond the SM \cite{blas}. The advantage of initial muon beam polarization is that it is effective for the indirect search \cite{fuku, ska}. Morever, it is versatile in the fundamental particle physics and nuclear physics \cite{gor, cohen}. In our previous work, investigation of the scalar unparticle and anomalous couplings at muon colliders in the RS model is considered in detail in Ref.\cite{gs2024}. However, the phenomenology of the $l^{-}l^{+}\nu \overline{\nu}$ final state at muon colliders  through the exclusive decay of ZZ/WW gauge bosons in the RS model
has not yet been invested. It is the worth that exclusive decay of the WW/ZZ gauge bosons into $l^{-}l^{+}\nu \overline{\nu}$ final state can be used to test these theoretical frameworks in a context where the energy scales are sufficiently large.\\
\hspace*{1cm} In this work, we calculate in detail the cross-sections for  the production of $l^{-}l^{+}\nu \overline{\nu}$ final state at multi-TeV  muon colliders through the exclusive decay of ZZ/WW gauge bosons in the RS model. The work is organized as follows. Theoretical framework is introduced in Section 2. The cross-sections for the production of the $l^{-}l^{+}\nu \overline{\nu}$ final state at muon colliders are calculated in detail in Section III. Finally, we summarize our results and make conclusions in Section IV. 
 \section{Theoretical framework}
\hspace*{1cm} In this section, we shortly review the popular RS scenario or RS1. The RS scenario consists of a 5D space compactified under a $S^{1}/Z^{2}$ orbifold symmetry \cite{rs, giori}. After compactification, one finds a 5D bulk space bounded by two branes. The RS model consider a non-factorizable 5-dimensional metric in the form 
\begin{equation}
d_{s}^{2} = e^{-2 kr_{c}|y|} \eta_{\mu \nu} d x^{\mu} d x^{\nu} - r_{c}^{2} d y^{2},
\end{equation}
where $k$ is the warping parameter of order the apparent gravitational scale $M_{P}$. The length-scale $r_{c}$ is related to the size of the extra-dimension. In particular, for $kr_{c} \sim 10$ the RS scenario can address the hierarchy problem. \\
\hspace*{1cm} Expanding the 4-dimensional component of the metric at first order about its static solution as follows \cite{folga}
\begin{equation}
G^{(5)}_{\mu\nu} = e^{-2r_{c}|y|} (\eta_{\mu\nu} + \kappa_{5}h_{\mu\nu}),
\end{equation}
with $\kappa_{5} = 2M_{5}^{-2/3}$, $M_{5}$ is the fundamental gravitational scale. The 5-dimensional field $h_{\mu\nu}$ can be written as a KK tower of 4-dimensional fields,
\begin{equation}
h_{\mu\nu} (x, y) = \sum_{n} h_{\mu\nu}^{n}(x) \frac{\chi^{n}(y)}{\sqrt{r_{c}}}.
\end{equation}
where the $h_{\mu\nu}^{n}(x)$ can be interpreted as the KK-excitations of the 4-dimensional graviton. The $\chi^{n}(y)$ factors are the wave-functions of the KK-gravitons along the extra dimension. The framework of a 5-dimensional metric yields several components upon 4-dimensional decomposition, including two noteworthy fields: the graviphoton ($G^{(5)}_{\mu 5}$) and the graviscalar ($G^{(5)}_{55}$). Due to the disruption of 5-dimensional translational invariance by the branes, the graviphoton is known to be massive. The graviscalar field, with the role in ensuring the stability of the extra-dimensional size, is introduced to be radion. The masses of the KK-graviton modes are given by
\begin{equation}
m_{n} = kx_{n}e^{-k\pi r_{c}}.
\end{equation}
\hspace*{1cm} The effective Lagrangian between KK-gravitons and brane matter is given by
\begin{equation}\label{eq1}
\mathcal{L} =  -\frac{1}{M_{5}^{3/2}} T^{\mu\nu} (x) h_{\mu\nu} (x, y = \pi) = -\frac{1}{M_{5}^{3/2}} T^{\mu\nu} (x)\sum_{n = 0} h_{\mu\nu}^{n}(x)\frac{\chi^{n}(y)}{\sqrt{r_{c}}}.\\
\end{equation}
For $n = 0$, the wave function at the IR brane location $y = \pi$ takes the form
\begin{equation}
\chi^{0} (y = \pi) = \sqrt{kr_{c}} \left(1 - e^{-2k\pi r_{c}}\right) = -\sqrt{r_{c}}\dfrac{M_{5}^{3/2}}{M_{P}},
\end{equation}
whereas for $n > 0$, 
\begin{equation}
\chi^{n} (y = \pi) = \sqrt{kr_{c}} e^{k\pi r_{c}} = \sqrt{r_{c}} e^{k\pi r_{c}} \dfrac{M_{5}^{3/2}}{M_{P}} = \sqrt{r_{c}}\dfrac{M_{5}^{3/2}}{\Lambda}.
\end{equation}
Therefore, Eq.(\ref{eq1}) becomes
\begin{equation}
\mathcal{L} = -\frac{1}{M_{P}}T^{\mu\nu} (x) h_{\mu\nu}^{0} (x) - \frac{1}{\Lambda}T^{\mu\nu}(x) h_{\mu\nu}^{n}(x).
\end{equation}
The $T^{\mu\nu}$ is the energy-momentum tensor, which is given at the tree level %\cite{csa}
\begin{equation}
T_{\mu \nu}=T_{\mu \nu}^{S M}+T_{\mu \nu}^{D M},
\end{equation}
where
\begin{equation}
\begin{aligned}
T_{\mu \nu}^{S M}= & {\left[\frac{i}{4} \bar{\psi}\left(\gamma_\mu D_\nu+\gamma_\nu D_\mu\right) \psi-\frac{i}{4}\left(D_\nu \bar{\psi} \gamma_\mu+D_\mu \bar{\psi} \gamma_\nu\right) \psi-\eta_{\mu \nu}\left(\bar{\psi} \gamma^\mu D_\mu \psi-m_\psi \bar{\psi} \psi\right)+\right.} \\
& \left.+\frac{i}{2} \eta_{\mu \nu} \partial^\rho \bar{\psi} \gamma_\rho \psi\right]+\left[\frac{1}{4} \eta_{\mu \nu} F^{\lambda \rho} F_{\lambda \rho}-F_{\mu \lambda} F_\nu^\lambda\right]+\left[\eta_{\mu \nu} D^\rho H^{\dagger} D_\rho H+\eta_{\mu \nu} V(H)+\right. \\
& \left.+D_\mu H^{\dagger} D_\nu H+D_\nu H^{\dagger} D_\mu H\right],
\end{aligned}
\end{equation}
and
\begin{equation}
T_{\mu \nu}^{D M}=\left(\partial_\mu S\right)\left(\partial_\nu S\right)-\frac{1}{2} \eta_{\mu \nu}\left(\partial^\rho S\right)\left(\partial_\rho S\right)+\frac{1}{2} \eta_{\mu \nu} m_S^2 S^2 + ...,
\end{equation}
\begin{equation}
T^{\mu}_{\mu}=\Sigma_{f} m_{f} \overline{f}f - 2m^{2}_{W}W^{+}_{\mu}W^{-\mu}-m^{2}_{Z}Z_{\mu}Z^{\mu} + (2m^{2}_{h_{0}}h_{0}^{2} -  \partial_{\mu}h_{0}\partial^{\mu}h_{0}) + ...
\end{equation}
\hspace*{1cm} The radion, as for the KK-graviton case, also couples with both the DM and SM particles through the trace of the energy-momentum tensor $T^{\mu}_{\mu}$ \cite{berna}. The radion Lagragian is given by the following form \cite{goldbe, csaki}:
\begin{equation}
\mathcal{L}_{\phi} = \frac{1}{2} (\partial_{\mu}\phi)(\partial^{\mu}\phi) - \frac{1}{2}m_{\phi}^{2}\phi^{2}+\frac{1}{\Lambda_{\phi}}\phi T^{\mu}_{\mu} + \frac{\alpha_{EM}C_{EM}}{8\pi\Lambda_{\phi}}\phi F_{\mu\nu}F^{\mu\nu} + \frac{\alpha_{S}C_{3}}{8\pi\Lambda_{\phi}}\phi\sum_{a}F^{a}_{\mu\nu}F^{a\mu\nu}.
\end{equation} 
where $F_{\mu\nu}$, $F^{a}_{\mu\nu}$ are the Maxwell and $SU_{c}(3)$ Yang-Mills tensors, respectively. The VEV of the radion $\Lambda_{\phi}$ is related to the effective scale $\Lambda$ such that $\Lambda_{\phi} = \sqrt{6}\Lambda$ \cite{park}. The $C_{3}$ and $C_{EM}$ are given by
\begin{align}
&C_{3} = b_{IR}^{(3)} - b_{UV}^{(3)} + \dfrac{1}{2}\sum_{q}F_{1/2}(x_{q}),\\
&C_{EM} = b_{IR}^{(EM)} - b_{UV}^{(EM)} + F_{1}(x_{W}) - \sum_{q}N_{C} Q_{q}^{2}F_{1/2}(x_{q}),
\end{align}
here $b_{IR}^{(3)} - b_{UV}^{(3)} = -11 + 2n/3$ where n is the number of quarks whose mass is smaller than $m_{\phi}/2$,  $b_{IR}^{(EM)} - b_{UV}^{(EM)} = 11/3$. The auxiliary functions $F_{1/2}, F_{1}$ are given by in Ref.\cite{blum}.\\
Feynman rules corresponding to the couplings of SM particles with KK-gravitons, radion, Higgs boson and scalar unparticle are given in Appendix B which is useful in our calculations.
  \section{The cross-section for the $l^{-}l^{+}\nu \overline{\nu}$ production in final state at multi-TeV muon colliders through the exclusive decay of WW/ZZ gauge bosons}
\hspace*{1cm} Influence of unparticle and polarization on properties of high energy colliders have been intensively studied in Refs.\cite{pra,alan,maj, kuma,sahi,kiku,chen,kha, fried, iltan, giang,soa1}.  Search for dark matter and unparticles produced in association with a Z boson in proton-proton collisions at $\sqrt{s} = 8$ TeV has been presented in Ref.\cite{khac}. The scalar unparticle signals at LHC with the center of mass energy as 14 TeV are investigated in Ref.\cite{alie}. Recently, investigation of the scalar unparticle and anomalous couplings at muon colliders in the RS model is considered in detail in Ref.\cite{gs2024}. A measurement of $l^{-}l^{+}\nu \overline{\nu}$ channel is considered in $\mu^{+}\mu^{-} \rightarrow W^{+} W^{-} \rightarrow l^{-}l^{+}\nu \overline{\nu}$ process in Ref.\cite{ZLu} and $p\overline{p}$ collision through the decay of $ ZZ \rightarrow l^{-}l^{+}\nu \overline{\nu}$ in Ref.\cite{victor}. In this section, we will evaluate the influence of scalar unparticle and polarization on exclusive decays of the WW/ZZ gauge bosons into $l^{-}l^{+}\nu \overline{\nu}$ final state at multi-TeV muon colliders in the RS model. Noting that the exclusive decay of the gauge bosons into final state containing a pair of leptons and neutrinos at high energy colliders can be used to test these theoretical frameworks in a context where the energy scales are sufficiently large. \\
\subsection{The transition amplitude for pair production of WW charged bosons }
\hspace*{1cm}First, we consider the collision process $\mu^{+}\mu^{-} \rightarrow W^{+}W^{-}$,  
\begin{equation} \label{pt4}
\mu^{-}(p_{1}) + \mu^{+}(p_{2}) \    \rightarrow         W^{-} (k_{1}) + W^{+} (k_{2}).
\end{equation}
\hspace*{1cm}In $\mu^{+}\mu^{-} \rightarrow W^{+}W^{-}$ process, the SM framework includes the tree level $\gamma, Z$ exchange in the s-channel and neutrino exchange in the t-channel. Using Feynman diagrams in Appendix A (Fig.\ref{Fig.6}), we can write the transition amplitudes for this process representing s-channel as follows
\begin{equation}
M_{s} = M_{\gamma} + M_{Z} + M_{\phi} + M_{h} + M_{U} + M_{G_{n}},
\end{equation}
where
%\begin{equation}
\begin{align}
&M_{\gamma} =  i \dfrac{e}{q^{2}_{s}}\overline{v}(p_{2})\gamma^{\sigma} u(p_{1}) \eta_{\sigma \beta} \varepsilon^{*}_{\mu} (k_{1}) \Gamma_{\gamma WW}^{\beta\mu\nu} \varepsilon^{*}_{\nu} (k_{2}),\\
&M_{Z} = -i \dfrac{g}{4 c_{W} (q^{2}_{s} - m^{2}_{Z})}\overline{v}(p_{2}) \gamma^{\sigma} \left(-1 +4s^{2}_{W} + \gamma_{5}\right) u(p_{1}) \left(\eta^{\sigma\beta} - \dfrac{q_{s}^{\sigma}q_{s}^{\beta}}{m^{2}_{Z}} \right) \varepsilon^{*}_{\mu} (k_{1})\Gamma_{ZWW}^{\beta\mu\nu} \varepsilon^{*}_{\nu} (k_{2}),\\
&M_{\phi} = -i \dfrac{\overline{g}_{\mu\mu\phi}\overline{g}_{W\phi}}{q^{2}_{s} - m^{2}_{\phi}}\overline{v}(p_{2})u(p_{1}) \varepsilon^{*}_{\mu} (k_{1}) \left[\eta^{\mu\nu} - 2g^{W}_{\phi}\left(\left(k_{1}k_{2}\right)\eta^{\mu\nu} - k_{1}^{\nu}k_{2}^{\mu}\right)\right]\varepsilon^{*}_{\nu} (k_{2}),\\
&M_{h} = - i\dfrac{\overline{g}_{\mu\mu h}\overline{g}_{Wh}}{q^{2}_{s} - m^{2}_{h}}\overline{v}(p_{2})u(p_{1}) \varepsilon^{*}_{\mu} (k_{1}) \left[\eta^{\mu\nu} - 2g^{W}_{h}\left(\left(k_{1}k_{2}\right)\eta^{\mu\nu} - k_{1}^{\nu}k_{2}^{\mu}\right)\right]\varepsilon^{*}_{\nu} (k_{2}),\\
&M_{U} = i\overline{g}_{\mu\mu U}\overline{g}_{WWU} \dfrac{A_{d_{U}}}{2sin(d_{U}\pi)} (-q^{2}_{s})^{d_{U} - 2}\overline{v}(p_{2})u(p_{1})\varepsilon^{*}_{\mu} (k_{1}) \left[\left(k_{1}k_{2}\right)\eta^{\mu\nu} - k_{1}^{\nu}k_{2}^{\mu}\right]\varepsilon^{*}_{\nu} (k_{2}),\\
&\begin{aligned}
M_{G_{n}} = - & \dfrac{i}{4\Lambda^{2}}\varepsilon^{*}_{\mu} (k_{1}) (m^{2}_{W} C_{\mu\nu\alpha\beta} + W_{\mu\nu\alpha\beta})\varepsilon^{*}_{\nu} (k_{2})\dfrac{P_{\sigma\rho\alpha\beta}}{q_{s}^{2}-m^{2}_{G_{n}} + i m_{G_{n}} \Gamma_{n}} \times\\
&  \overline{v}(p_{2}) \left[\gamma_{\sigma}(p_{1\rho} - p_{2\rho}) + \gamma_{\rho}(p_{1\sigma} - p_{2\sigma}) - 2 \eta_{\sigma\rho} (\widehat{p_{1}} - \widehat{p_{2}} - 2m_{\mu}) \right].
\end{aligned}
\end{align}
\hspace*{1cm}The transition amplitude representing t-channel can be written as
\begin{equation}
M_{t} = -\frac{g^{2}}{2 (q^{2}_{t} - m^{2}_{\nu})} \varepsilon^{*}_{\nu} (k_{2}) \overline{v}(p_{2}) \gamma^{\mu} \frac{1 - \gamma^{5}}{2} \widehat{q}_{t} \varepsilon^{*}_{\mu} (k_{1}) \gamma^{\nu} \frac{1 - \gamma^{5}}{2} u(p_{1}).
\end{equation}
Here, $q_{s} = p_{1} + p_{2} = k_{1} + k_{2}$, $s = (p_{1} + p_{2})^{2}$ is the square of the collision energy, $M_{U}$ is the contribution by the scalar unparticle, which is important for the described process. $\overline{g}_{\mu\mu\phi}, \overline{g}_{W\phi}, \overline{g}_{\mu\mu h}, \overline{g}_{Wh}$ are couplings shown in Ref.\cite{ahm}. $\Gamma_{\gamma WW}^{\beta\mu\nu}, \Gamma_{ZWW}^{\beta\mu\nu} $ tensors are given by \cite{bha}. $C_{\mu\nu\alpha\beta}$, $ W_{\mu\nu\alpha\beta}$, $P_{\sigma\rho\alpha\beta}$, $\Gamma_{n}$ are given by \cite{folga}.
\subsection{The transition amplitude for pair production of the ZZ neutral bosons}
\hspace*{1cm}Next, we consider the collision process $\mu^{-}\mu^{+} \rightarrow ZZ$
\begin{equation}
\mu^{-}(p_{1}) + \mu^{+}(p_{2}) \    \rightarrow       Z (k_{1}) + Z (k_{2}).
\end{equation}
\hspace*{1cm}In the SM, the process $\mu^{-}\mu^{+} \rightarrow ZZ$ proceeds via the u-, t-channel exchange diagrams at tree level. The presence of new physics is via s-channel. The Feynman diagrams are shown in Appendix A (Fig.\ref{Fig.7}).\\
\hspace*{1cm}The transition amplitude representing the s-channel is given by
\begin{equation}
M_{s} = M_{Z} + M_{\gamma} + M_{h} + M_{\phi} + M_{U} + M_{G_{n}},
\end{equation}where 
\begin{align}
&M_{Z} = \dfrac{-i g}{4 c_{W} (q^{2}_{s} - m^{2}_{Z})} \varepsilon^{*}_{\mu} (k_{1}) \Gamma^{\sigma\mu\nu}_{ZZZ}(q_{s}, k_{1}, k_{2})\varepsilon^{*}_{\nu} (k_{2})\left(\eta_{\sigma\beta} - \dfrac{q_{s\sigma}q_{s\beta}}{m^{2}_{Z}}\right)\overline{v}(p_{2}) \gamma^{\beta}\left(-1 + 4s^{2}_{W} + \gamma^{5}\right) u(p_{1}),\\
&M_{\gamma} = \dfrac{-e}{q^{2}_{s}} \varepsilon^{*}_{\mu} (k_{1}) \Gamma^{\sigma\mu\nu}_{\gamma ZZ}(q_{s}, k_{1}, k_{2})\varepsilon^{*}_{\nu} (k_{2})\eta_{\sigma\beta} \overline{v}(p_{2}) \gamma^{\beta} u(p_{1}),\\
&M_{h} = \dfrac{\overline{g}_{\mu\mu h}\overline{g}_{hZ}}{q^{2}_{s} - m^{2}_{h}}\varepsilon^{*}_{\mu} (k_{1}) \left[\eta^{\mu\nu} - 2g^{Z}_{h}\left(\left(k_{1}k_{2}\right)\eta^{\mu\nu} - k_{1}^{\nu}k_{2}^{\mu}\right)\right]\varepsilon^{*}_{\nu} (k_{2})\overline{v}(p_{2})u(p_{1}),\\
&M_{\phi} = \dfrac{\overline{g}_{\mu\mu\phi}\overline{g}_{\phi Z}}{q^{2}_{s} - m^{2}_{\phi}}\varepsilon^{*}_{\mu} (k_{1}) \left[\eta^{\mu\nu} - 2g^{Z}_{\phi}\left(\left(k_{1}k_{2}\right)\eta^{\mu\nu} - k_{1}^{\nu}k_{2}^{\mu}\right)\right]\varepsilon^{*}_{\nu} (k_{2})\overline{v}(p_{2})u(p_{1}),\\
&M_{U} =  -i\overline{g}_{\mu\mu U}\overline{g}_{ZZU} \dfrac{A_{d_{U}}}{2sin(d_{U}\pi)} (-q^{2}_{s})^{d_{U} - 2}\varepsilon^{*}_{\mu} (k_{1})\left[\left(k_{1}k_{2}\right)\eta^{\mu\nu} - k_{1}^{\nu}k_{2}^{\mu}\right]\varepsilon^{*}_{\nu} (k_{2})\overline{v}(p_{2})u(p_{1}),\\
&\begin{aligned}
M_{G_{n}} = - & \dfrac{i}{4\Lambda^{2}}\varepsilon^{*}_{\mu} (k_{1}) (m^{2}_{Z} C_{\mu\nu\alpha\beta} + W_{\mu\nu\alpha\beta})\varepsilon^{*}_{\nu} (k_{2})\dfrac{P_{\sigma\rho\alpha\beta}}{q_{s}^{2}-m^{2}_{G_{n}} + i m_{G_{n}} \Gamma_{n}} \times\\
&  \overline{v}(p_{2}) \left[\gamma_{\sigma}(p_{1\rho} - p_{2\rho}) + \gamma_{\rho}(p_{1\sigma} - p_{2\sigma}) - 2 \eta_{\sigma\rho} (\widehat{p_{1}} - \widehat{p_{2}} - 2m_{\mu}) \right].
\end{aligned}
\end{align}
\hspace*{1cm}The transition amplitude representing the u-channel is given by
\begin{equation}
\begin{aligned}
M_{u} = &-i\dfrac{g^{2}}{16c^{2}_{W}(q^{2}_{u} - m^{2}_{\mu})}\overline{v}(p_{2})\gamma^{\mu}\left(-1 + 4s^{2}_{W} + \gamma^{5}\right)\varepsilon^{*}_{\mu} (k_{1})\times\\
&\left(\slashed{q}_{u}+m_{\mu}\right)\gamma^{\nu}\left(-1 + 4s^{2}_{W} + \gamma^{5}\right) \varepsilon^{*}_{\nu} (k_{2})u(p_{1}).
\end{aligned}
\end{equation}
\hspace*{1cm}The transition amplitude representing the t-channel is given by
\begin{equation}
\begin{aligned}
M_{t} = &-i\dfrac{g^{2}}{16c^{2}_{W}(q^{2}_{t} - m^{2}_{\mu})}\overline{v}(p_{2})\gamma^{\nu}\left(-1 + 4s^{2}_{W} + \gamma^{5}\right)\varepsilon^{*}_{\nu} (k_{2})\times\\
&\left(\slashed{q}_{t}+m_{\mu}\right)\gamma^{\mu}\left(-1 + 4s^{2}_{W} + \gamma^{5}\right) \varepsilon^{*}_{\mu} (k_{1})u(p_{1}).
\end{aligned}
\end{equation}
Here, $q_{s} = p_{1} + p_{2} = k_{1} + k_{2}$, $q_{u} = p_{1} - k_{2} = k_{1} - p_{2}$, $q_{t} = p_{1} - k_{1} = k_{2} - p_{2}$, $s = (p_{1} + p_{2})^{2}$ is the square of the collision energy. $\overline{g}_{\mu\mu\phi}, \overline{g}_{\phi Z}, \overline{g}_{\mu\mu h}, \overline{g}_{hZ}$ are couplings shown in Ref.\cite{ahm}. The triple gauge boson couplings $\Gamma^{\sigma\mu\nu}_{\gamma ZZ}$, $\Gamma^{\sigma\mu\nu}_{ZZZ}$ are given by \cite{raha}. $C_{\mu\nu\alpha\beta}$, $ W_{\mu\nu\alpha\beta}$, $P_{\sigma\rho\alpha\beta}$, $\Gamma_{n}$ are given by \cite{folga}.
\subsection{Numerical evaluation for cross-sections of the $l^{-}l^{+}\nu \overline{\nu}$ final state}
\hspace*{1cm} In general, theoretical cross-section depends on the polarization as follows \cite{spo}
\begin{equation}
\sigma = \dfrac{1 - P_{\mu^{-}}}{2}\dfrac{1 - P_{\mu^{+}}}{2}\sigma_{LL} + \dfrac{1 + P_{\mu^{-}}}{2}\dfrac{1 + P_{\mu^{+}}}{2}\sigma_{RR} + \dfrac{1 - P_{\mu^{-}}}{2}\dfrac{1 + P_{\mu^{+}}}{2}\sigma_{LR} + \dfrac{1 + P_{\mu^{-}}}{2}\dfrac{1 - P_{\mu^{+}}}{2}\sigma_{RL},
\end{equation}
where $\sigma_{ij} (ij = LL, RR, LR, RL)$ are the cross-sections with the $"i, j"$ standing for the polarization of the $\mu^{-}, \mu^{+}$ beams (left or right-handed), respectively. \\
 \hspace*{1cm} The total cross-section for the $l^{-}l^{+}\nu \overline{\nu}$ final states can be calculated as follows
\begin{equation} \label{sigmaW}
\sigma_{l^{-}l^{+}\nu \overline{\nu}} = \sigma (\mu^{-} \mu^{+} \rightarrow  W^{+}W^{-}) \times   Br(W^{-}\rightarrow l^{-}\overline{\nu}_{l}) Br(W^{+} \rightarrow l^{+}\nu_{l})
\end{equation}
and
\begin{equation} \label{sigmaZ}
\sigma_{l^{-}l^{+}\nu \overline{\nu}} = \sigma (\mu^{-} \mu^{+} \rightarrow  Z_1Z_2) \times   Br(Z_1\rightarrow l^{-}l^{+}) Br(Z_2 \rightarrow \nu_{l}\overline{\nu}_{l}).
\end{equation}
 \hspace*{1cm} Here, $\sigma (\mu^{-} \mu^{+} \rightarrow  W^{+}W^{-}/Z_1Z_2)$ can be calculated from the expressions of the differential cross-section \cite{pes}
\begin{equation}
\frac{d\sigma (\mu^{-} \mu^{+} \rightarrow  W^{+}W^{-}/Z_1Z_2)}{dcos\psi} = \frac{1}{32 \pi s} \frac{|\overrightarrow{k}_{1}|}{|\overrightarrow{p}_{1}|} |M_{fi}|^{2},
\end{equation}
where $\psi = (\overrightarrow{p}_{1}, \overrightarrow{k}_{1})$ is the scattering angle. For the term $|M_{fi}|^{2}$, we have taken the average over initial, the sum over final spins and the polarization sum of gauge bosons (W/Z). The model parameters are chosen as $\xi=  1/6$ \cite{ahm}, $\lambda_{\mu \mu} = \lambda_{WW} = \lambda_{ZZ} = \lambda_{0} = 1$ \cite{khacha}, $ m_{h}$ = 125 GeV, $m_{\phi}$ = 125 GeV \cite{giang2023}. Based on the results in Ref.\cite{khacha}, the unparticle scale $\Lambda_{U}$ in the $1-5$ TeV range is chosen because it is high enough to satisfy many bounds yet low enough to give observable collider effects. With $k/M_{P} \leq 0.1$, the first KK-graviton excitation becomes the primary observable signature to elucidate the fundamental parameters of RS \cite{davo}. The present experimental bounds on the parameters of the model as follows: the mass of the first KK-graviton $m_{G_{1}} = 1$ TeV, the effective scale $\Lambda = 10$ TeV \cite{folga1, mgfol}. The integrated luminosity scaling of a high energy muon collider (assuming a 5 year run) reaches 1 $ab^{-1}$ (3 TeV), 4 $ab^{-1}$ (6 TeV), 10 $ab^{-1}$ (10 TeV) and 20 $ab^{-1}$ (14 TeV) \cite{asadi, cap, spor, liu}. In this work, we choose a high energy muon collider 10 TeV using 10 $ab^{-1}$ of integrated luminosity. The bounds on the anomalous $W^{-}W^{+}\gamma$ and $W^{-}W^{+}Z$ couplings are provided by the LEP, Tevatron and LHC experiments. The ATLAS collaboration has updated the best available constraints on anomalous couplings $\Delta k_{\gamma}$, $\lambda_{\gamma}$, $\Delta k_{Z}$, $\lambda_{Z}$, $f^{\gamma}_{4}$, $f^{Z}_{4}$, $f^{\gamma}_{5}$, $f^{Z}_{5}$ obtained as follows: $\Delta k_{\gamma} \in [-0.135, 0.190]$, $\lambda_{\gamma} \in [-0.065, 0.061]$, $\Delta k_{Z} \in [-0.061, 0.093]$, $\lambda_{Z}\in [-0.062, 0.065]$ \cite{cakir}, $f^{\gamma}_{4} \in [-0.0012, 0.0012]$, $f^{Z}_{4} \in [-0.001, 0.001]$, $f^{\gamma}_{5} \in [-0.0012, 0.0012]$, $f^{Z}_{5} \in [-0.001, 0.001]$ \cite{navas}. From the formulas (\ref{sigmaW} - \ref{sigmaZ}) with a branching ratio in agreement with the SM prediction \cite{navas}, we give estimates in detail for the cross-sections as follows:\\

\begin{figure}[!htb] 
\begin{center}
    \begin{tabular}{cc}
        \includegraphics[width=8cm, height= 5cm]{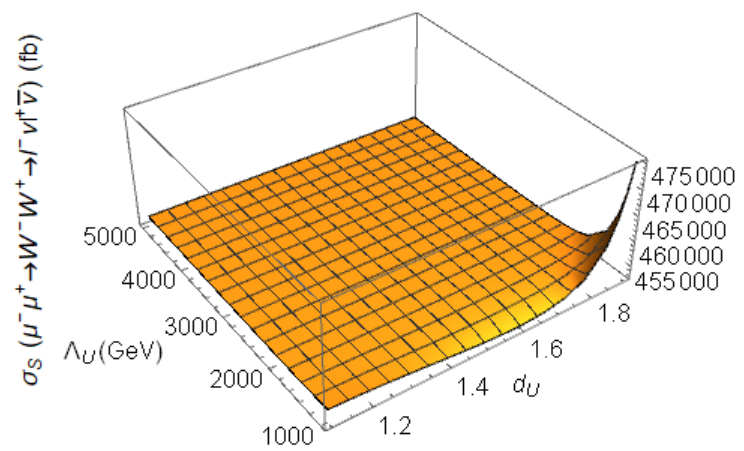} &
        \includegraphics[width=8cm, height= 5cm]{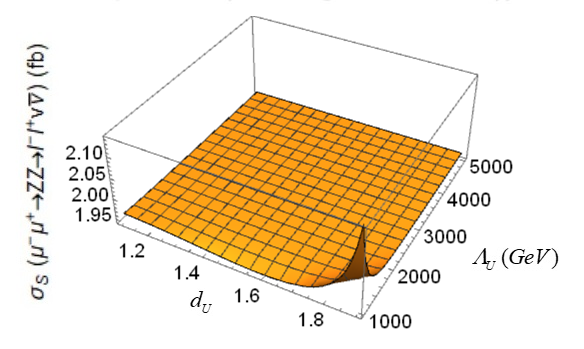}\\
         \small(a)& \small(b)\\
             \end{tabular}
    \caption{\label{Fig.1}  The total cross-section depends on the ($\Lambda_{U}, d_{U}$) in (a) $\mu^{+}\mu^{-} \rightarrow W^{+}W^{-} \rightarrow l^{-}l^{+}\nu \overline{\nu}$, (b) $\mu^{+}\mu^{-} \rightarrow ZZ \rightarrow l^{-}l^{+}\nu \overline{\nu}$ collisions. The parameters are chosen as $\sqrt{s} = 10$ TeV, $P_{\mu^{-}} = 0.8, P_{\mu^{+}} = -0.8$,  $\Delta k_{\gamma} = -0.135$, $\lambda_{\gamma} = 0.061$, $\Delta k_{Z} = 0.093$, $\lambda_{Z} = -0.062$, $f^{\gamma}_{4} = -0.0012$, $f^{Z}_{4} = 0.001$, $f^{\gamma}_{5} = -0.0012$, $f^{Z}_{5} = 0.001$.}
    \end{center}
\end{figure}
\hspace*{0.5cm} i) We evaluate the benchmark background $(\Lambda_{U}, d_{U})$ in Fig.\ref{Fig.1}. The parameters are chosen as $P_{\mu^{-}} = 0.8, P_{\mu^{+}} = -0.8$, $\sqrt{s} = 10$ TeV, $\Delta k_{\gamma} = -0.135$, $\lambda_{\gamma} = 0.061$, $\Delta k_{Z} = 0.093$, $\lambda_{Z} = -0.062$, $f^{\gamma}_{4} = -0.0012$, $f^{Z}_{4} = 0.001$, $f^{\gamma}_{5} = -0.0012$, $f^{Z}_{5} = 0.001$. From the figures, we can see that the cross-sections for $\mu^{+}\mu^{-} \rightarrow W^{+}W^{-}/ZZ \rightarrow l^{-}l^{+}\nu \overline{\nu}$ collisions reach the maximum value at $(\Lambda_{U}, d_{U})$ $= (1 TeV, 1.9)$.  \\
\begin{figure}[!htb] 
\begin{center}
    \begin{tabular}{cc}
        \includegraphics[width=7.5cm, height= 4.5cm]{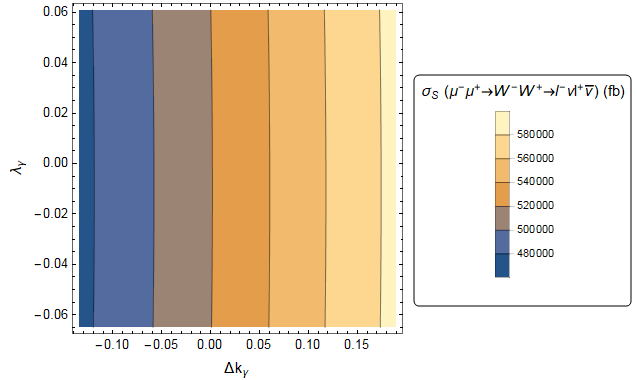} &
        \includegraphics[width=7.5cm, height= 4.5cm]{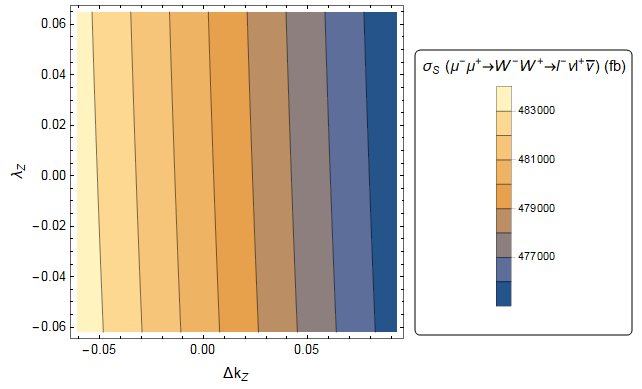} \\
         \small(a)& \small(b)\\
                    \end{tabular}
    \caption{\label{Fig.2}  The total cross-section in $\mu^{+}\mu^{-} \rightarrow W^{+}W^{-} \rightarrow l^{-}l^{+}\nu \overline{\nu}$ collision depends on the (a) ($\Delta k_{\gamma}, \lambda_{\gamma}$), (b) ($\Delta k_{Z}, \lambda_{Z}$). The parameters are chosen as $\sqrt{s} = 10$ TeV, $P_{\mu^{-}} = 0.8, P_{\mu^{+}} = -0.8$, $\Lambda_{U} = 1$ TeV, $d_{U} = 1.9$.}
    \end{center}
\end{figure}
\hspace*{0.5cm} ii) The total cross - section for  $\mu^{+}\mu^{-} \rightarrow W^{+}W^{-} \rightarrow l^{-}l^{+}\nu \overline{\nu}$ collision depends on ($\Delta k_{\gamma}$, $\lambda_{\gamma}$) shown in the Fig.\ref{Fig.2}a. The parameters are taken to be $P_{\mu^{-}} = 0.8, P_{\mu^{+}} = -0.8$, $\sqrt{s} = 10$ TeV, $\Lambda_{U} = 1$ TeV, $d_{U} = 1.9$, $\Delta k_{Z} = 0.093$, $\lambda_{Z} = -0.062$.  With the fixed value of $\Delta k_{\gamma}$, the cross-section is independent on $\lambda_{\gamma}$ values. The result shows that cross-section is the largest in the yellow region of the figure in which typical value is given by $\sigma \approx 58\times 10^4$ fb for the final state. The total cross - section depends on ($\Delta k_{Z}$, $\lambda_{Z}$) shown in the Fig.\ref{Fig.2}b. With the fixed value of ($\Delta k_{\gamma}$, $\lambda_{\gamma}$) $= (-0.135, 0.061)$, the largest value of cross-section is about $48.3\times 10^4$ fb.\\
\begin{figure}[!htb] 
\begin{center}
   \begin{tabular}{cc}
      \includegraphics[width=7.5cm, height= 4.5cm]{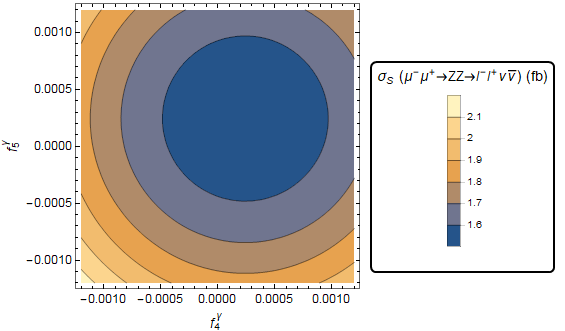} &
     \includegraphics[width=7.5cm, height= 4.5cm]{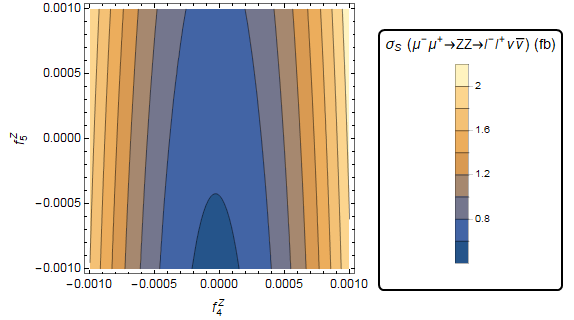} \\
           \small(a)& \small(b)\\
            \end{tabular}
    \caption{\label{Fig.3}  The total cross-section in $\mu^{+}\mu^{-} \rightarrow ZZ \rightarrow l^{-}l^{+}\nu \overline{\nu}$ collision depends on the (a) ($f^{\gamma}_{4}, f^{\gamma}_{5}$), (b) ($f^{Z}_{4}, f^{Z}_{5}$). The parameters are chosen as $\sqrt{s} = 10$ TeV, $P_{\mu^{-}} = 0.8, P_{\mu^{+}} = -0.8$, $\Lambda_{U} = 1$ TeV, $d_{U} = 1.9$.}
    \end{center}
\end{figure}
\hspace*{0.5cm} iii)  With the parameters are chosen as above, i.e  $(\Lambda_{U}, d_{U}) = (1 TeV, 1.9)$, $P_{\mu^{-}} = 0.8, P_{\mu^{+}} = -0.8$ and $\sqrt{s} = 10$ TeV, the total cross - section for $\mu^{+}\mu^{-} \rightarrow ZZ \rightarrow l^{-}l^{+}\nu \overline{\nu}$ collision depends on the parameters of ($f^{\gamma}_{4}$, $f^{\gamma}_{5}$) and ($f^{Z}_{4}$, $f^{Z}_{5}$) . With the fixed value  of $(f^{Z}_{4}, f^{Z}_{5}) = (0.001, 0.001)$, the total cross - section depends on the parameters ($f^{\gamma}_{4}$, $f^{\gamma}_{5}$) shown in  the Fig.\ref{Fig.3}a and with $(f^{\gamma}_{4}, f^{\gamma}_{5}) = (-0.0012, -0.0012)$,  the total cross - section depends on the parameters ($f^{Z}_{4}, f^{Z}_{5}$) shown in Fig.\ref{Fig.3}b. We can see from the figures that largest values of cross-section are given by,  $\sigma \approx 2.1$ fb for Fig.\ref{Fig.3}a and $\sigma\approx  2$ fb  for Fig.\ref{Fig.3}b,  respectively. \\
\begin{figure}[!htb] 
\begin{center}
 \begin{tabular}{cc}
            \includegraphics[width=8cm, height=5cm]{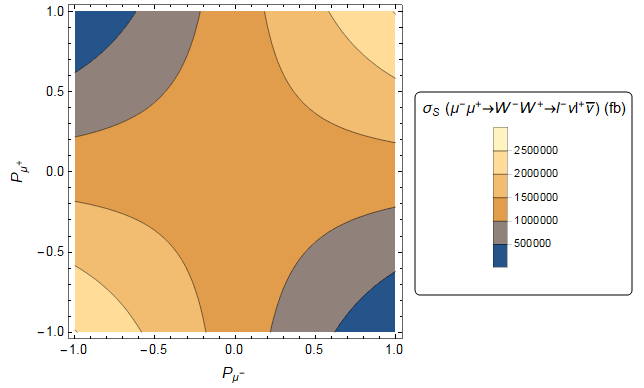} &
        \includegraphics[width=8cm, height= 5cm]{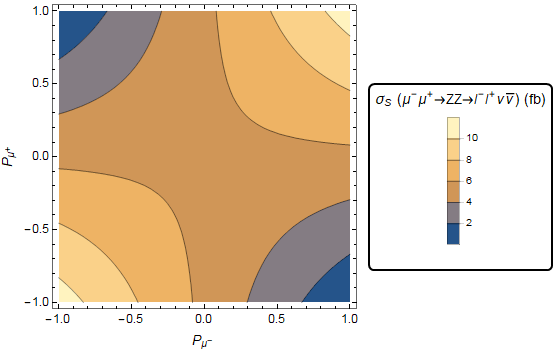} \\
         \small(a)& \small(b)
          \end{tabular}
            \caption{\label{Fig.4} The total cross-section as a function of the polarization coefficients of muon and antimuon beam in (a) $\mu^{+}\mu^{-} \rightarrow W^{+}W^{-} \rightarrow l^{-}l^{+}\nu \overline{\nu}$, (b) $\mu^{+}\mu^{-} \rightarrow ZZ \rightarrow l^{-}l^{+}\nu \overline{\nu}$ collisions. The parameters are chosen as $\sqrt{s} = 10$ TeV, $\Lambda_{U} = 1$ TeV, $d_{U} = 1.9$, $\Delta k_{\gamma} = -0.135$, $\lambda_{\gamma} = 0.061$, $\Delta k_{Z} = 0.093$, $\lambda_{Z} = -0.062$, $f^{\gamma}_{4} = -0.0012$, $f^{Z}_{4} = 0.001$, $f^{\gamma}_{5} = -0.0012$, $f^{Z}_{5} = 0.001$.}
    \end{center}
\end{figure}
 \hspace*{0.5cm} iv)  With the parameters chosen as in Fig.\ref{Fig.1}, the total cross-sections as the function of the polarization coefficients ($P_{\mu^{-}},P_{\mu^{-}}$) are plotted in Fig.\ref{Fig.4}. The figure indicates that the total cross-section for $\mu^{+}\mu^{-} \rightarrow W^{+}W^{-}/ZZ \rightarrow l^{-}l^{+}\nu \overline{\nu}$ collisions achieves the maximum value when $P_{\mu^{-}} = P_{\mu^{+}} = \pm 1$ and the minimum value when $P_{\mu^{-}} = 1, P_{\mu^{+}} = -1$ or $P_{\mu^{-}} = -1, P_{\mu^{+}} = 1$, respectively.\\
 \begin{figure}[!htb] 
\begin{center}
    \begin{tabular}{cc}
        \includegraphics[width=7.5cm, height= 5cm]{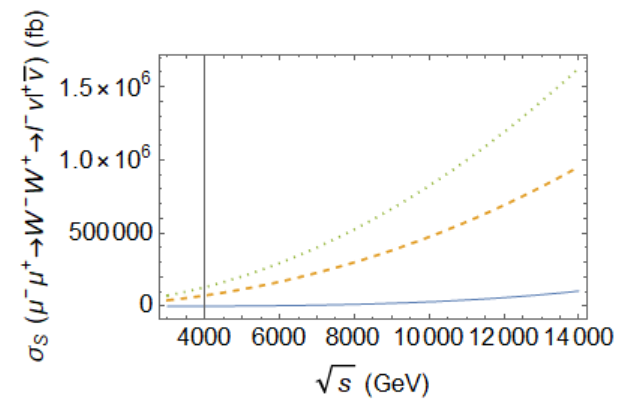} &
        \includegraphics[width=9cm, height= 4.5cm]{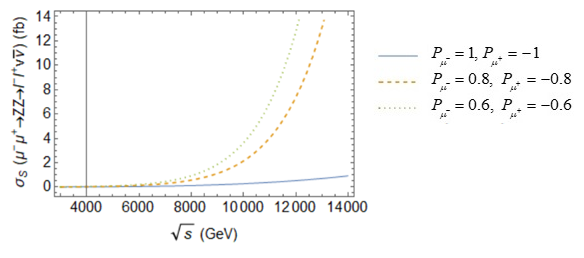}\\
         \small(a)& \small(b) \\
             \end{tabular}
    \caption{\label{Fig.5}  The total cross-section depends on the collision energy in (a) $\mu^{+}\mu^{-} \rightarrow W^{+}W^{-} \rightarrow l^{-}l^{+}\nu \overline{\nu}$, (b) $\mu^{+}\mu^{-} \rightarrow ZZ \rightarrow l^{-}l^{+}\nu \overline{\nu}$ collisions in case of $(P_{\mu^{-}}, P_{\mu^{+}}) = (1, -1), (0.8, -0.8), (0.6, -0.6)$. The parameters are chosen as $\Lambda_{U} = 1$ TeV, $d_{U} = 1.9$, $\Delta k_{\gamma} = -0.135$, $\lambda_{\gamma} = 0.061$, $\Delta k_{Z} = 0.093$, $\lambda_{Z} = -0.062$, $f^{\gamma}_{4} = -0.0012$, $f^{Z}_{4} = 0.001$, $f^{\gamma}_{5} = -0.0012$, $f^{Z}_{5} = 0.001$.}
    \end{center}
\end{figure}
\hspace*{0.5cm} v) With the model parameters chosen as in Fig.\ref{Fig.4} and the polarization coefficients taken from Ref.\cite{ZLu}, i.e  $(P_{\mu^{-}}, P_{\mu^{+}})$ $= (1, -1), (0.8, -0.8), (0.6, -0.6)$, the dependence of the total cross-section on the collision energy is shown in Fig.\ref{Fig.5}. From the figures, we can see that the cross-sections increase as the collision energy increases. The cross-sections for $\mu^{+}\mu^{-} \rightarrow W^{+}W^{-} \rightarrow l^{-}l^{+}\nu \overline{\nu}$ collision are about $10^{6}$ times larger than that of $\mu^{+}\mu^{-} \rightarrow ZZ \rightarrow l^{-}l^{+}\nu \overline{\nu}$ collision due to branching ratio of decay channels. It is worth that with the contribution of scalar unparticle and KK-graviton in s-channel, the cross-section for  $\mu^{+}\mu^{-} \rightarrow W^{+}W^{-} \rightarrow l^{-}l^{+} \nu \overline{\nu}$ collisions is greatly larger than that of  the  $Z^{'}$ heavy neutral gauge boson under the same conditions in Ref. \cite{ZLu}. \\ 

\begin{figure}[!htb] 
\begin{center}
    \begin{tabular}{cc}
        \includegraphics[width=8cm, height= 5cm]{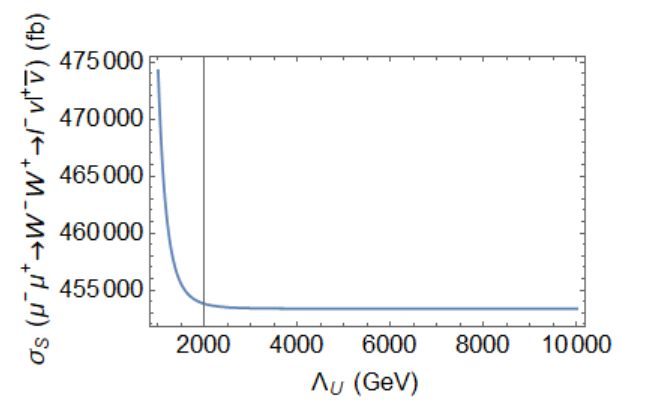} &
        \includegraphics[width=7.5cm, height= 4.5cm]{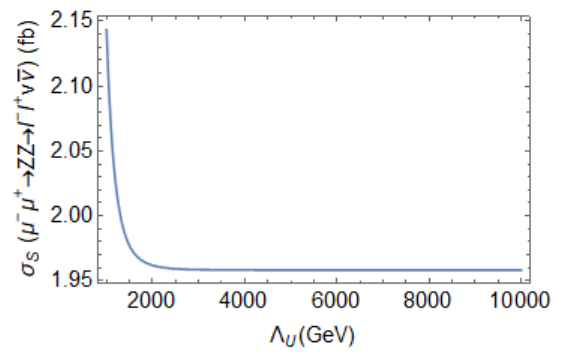}\\
         \small(a)& \small(b)\\
             \end{tabular}
    \caption{\label{Fig.8}  The total cross-section depends on the $\Lambda_{U}$ in (a) $\mu^{+}\mu^{-} \rightarrow W^{+}W^{-} \rightarrow l^{-}l^{+}\nu \overline{\nu}$, (b) $\mu^{+}\mu^{-} \rightarrow ZZ \rightarrow l^{-}l^{+}\nu \overline{\nu}$ collisions. The parameters are chosen as $d_{U} = 1.9$, $P_{\mu^{-}} = 0.8, P_{\mu^{+}} = -0.8$, $\sqrt{s} = 10$ TeV, $\Delta k_{\gamma} = -0.135$, $\lambda_{\gamma} = 0.061$, $\Delta k_{Z} = 0.093$, $\lambda_{Z} = -0.062$, $f^{\gamma}_{4} = -0.0012$, $f^{Z}_{4} = 0.001$, $f^{\gamma}_{5} = -0.0012$, $f^{Z}_{5} = 0.001$.}
    \end{center}
\end{figure}
   \hspace*{0.5cm} vi) With the parameters chosen as in Fig.\ref{Fig.1} and $d_{U}$ = 1.9, the total cross-sections depend on the parameter $\Lambda_{U}$  in the range of [1, 10] TeV are shown in Fig.\ref{Fig.8}. From the figures, we can see that the cross-sections decrease rapidly as the $\Lambda_{U}$ increases from 1 TeV to 2 TeV and then the shape of cross-section becomes flat as $\Lambda_{U}$ increases. It is interesting to note that with the bounds for $\Lambda_{U}$ are around of 1TeV, the unparticle signals are expected to be observed. This result is similar to Ref.\cite{alie}.\\
 \begin{table}[!htb]
 	  \caption{\label{tab1} The different values of the cross-section for $\mu^{+}\mu^{-} \rightarrow W^{+}W^{-} \rightarrow l^{-}\nu l^{+}\overline{\nu} $ collision correspond to the contribution of scalar particles ($\phi$, h, U), KK-graviton and SM propagators in case of $P_{\mu^{-}} = 0.8, P_{\mu^{+}} = -0.8$, $\Delta k_{\gamma} = -0.135$, $\lambda_{\gamma} = 0.061$, $\Delta k_{Z} = 0.093$, $\lambda_{Z} = -0.062$, $f^{\gamma}_{4} = -0.0012$, $f^{Z}_{4} = 0.001$, $f^{\gamma}_{5} = -0.0012$, $f^{Z}_{5} = 0.001$.} 
 	  	  \vspace*{0.5cm}
 	  	\begin{center}
 	  	  \begin{tabular}{|c|c|c|c|c|c|}     	 
 	 	  \hline
 	   $\sqrt{s}$ (TeV) & $\sigma_{\phi, h}$ (fb) & $\sigma_{G}$ (fb) & $\sigma_{Unparticle}$ (fb) & $\sigma_{SM} (fb)$ & $\sigma_{RS+U+anomalous} (fb)$\\
	\hline
	3 & 0.000226032 &  0.316343 & 330.404 & 41605.9 &  41819.2\\
	6 & 0.000226166 & 4.25979 & 4019.25 & 165792.3 & 168256.4\\
	10 & 0.000226194 & 31.7527 & 25298.6  &  460855.8 & 475433.5 \\
	14 & 0.000226202 & 120.933 & 84964.9 & 907644.6 & 954733.0\\
		\hline
		 \end{tabular}
    \end{center}
   \end{table}

     \begin{table}[!htb]
 	  \caption{\label{tab2}  The different values of the cross-section for  $\mu^{+}\mu^{-} \rightarrow ZZ \rightarrow l^{-}l^{+}\nu\overline{\nu} $ collision correspond to the contribution of scalar particles ($\phi$, h, U), KK-graviton, and SM propagators in case of $P_{\mu^{-}} = 0.8, P_{\mu^{+}} = -0.8$, $\Delta k_{\gamma} = -0.135$, $\lambda_{\gamma} = 0.061$, $\Delta k_{Z} = 0.093$, $\lambda_{Z} = -0.062$, $f^{\gamma}_{4} = -0.0012$, $f^{Z}_{4} = 0.001$, $f^{\gamma}_{5} = -0.0012$, $f^{Z}_{5} = 0.001$.} 
 	  	  \vspace*{0.5cm}
 	  	\begin{center}
 	  	  \begin{tabular}{|c|c|c|c|c|c|}     	 
 	 	  \hline
 	   $\sqrt{s}$ (TeV) & $\sigma_{\phi, h}$ ($ 10^{-8}$ fb) & $\sigma_{G}$ ($ 10^{-3}$ fb) & $\sigma_{Unparticle}$ (fb) & $\sigma_{SM} (fb)$ & $\sigma_{RS+U+anomalous} (fb)$\\
	\hline 
	3 & 0.19861  &0.00214 & 0.0029 & 0.00038 & 0.0083\\
	6 & 0.19892  & 0.02883  & 0.0354 & 0.00153 & 0.1366\\
	10 &  0.19899 & 0.2149   & 0.2226  & 0.00423 & 2.1493\\
	14 & 0.19901 & 0.81909  & 0.7475  & 0.00829 & 22.6408\\
		\hline
			 \end{tabular}
    \end{center}
   \end{table} 

\hspace*{0.5cm} vii)  With the parameters chosen as in Fig.\ref{Fig.1} and the benchmark signal point $(\Lambda_{U}, d_{U})$ = (1 TeV, 1.9), the different values of the cross-section correspond to the contribution of scalar particles ($\phi$, h, U), KK-graviton propagators and also the SM are given in Table.\ref{tab1} and Table.\ref{tab2}, respectively. From these tables  we can see that the contribution of the scalar unparticle in s-channel is much larger than that of the radion, Higgs and also KK-graviton under the same conditions. With the contributions of new physics, the total cross-section is greatly enhanced. For more details, we evaluate  the number of events (assuming in a 5 year run) with the different luminosity in Table.3. We can see from Table.3 that the number of events for $l^{-}l^{+}\nu\overline{\nu}$ final state through the exclusive decay of WW charged bosons are much larger than that of $ZZ$ neutral bosons which can  be measured in experiments.\\        
   \begin{table}[!htb]
 	  \caption{\label{tab3} Number of events (assuming in a 5 year run) in $\mu^{+}\mu^{-} \rightarrow W^{+}W^{-}/ZZ \rightarrow l^{-}l^{+}\nu\overline{\nu}$ collisions in case of $P_{\mu^{-}} = 0.8, P_{\mu^{+}} = -0.8$, $\Delta k_{\gamma} = -0.135$, $\lambda_{\gamma} = 0.061$, $\Delta k_{Z} = 0.093$, $\lambda_{Z} = -0.062$, $f^{\gamma}_{4} = -0.0012$, $f^{Z}_{4} = 0.001$, $f^{\gamma}_{5} = -0.0012$, $f^{Z}_{5} = 0.001$.} 
 	  	  \vspace*{0.5cm}
 	  	\begin{center}
 	  	  \begin{tabular}{|c|c|c|}     	 
 	 	  \hline
 	   $\mathcal{L}$ & $N(\mu^{+}\mu^{-} \rightarrow W^{+}W^{-} \rightarrow l^{-}\nu l^{+}\overline{\nu}) $ & $N(\mu^{+}\mu^{-} \rightarrow ZZ \rightarrow l^{-}l^{+}\nu\overline{\nu})$ \\
	\hline
	1 $ab^{-1}$ (3 TeV) & 4.1819 $\times 10^{7}$ &  8 \\
	4 $ab^{-1}$(6 TeV) & 6.7303 $\times 10^{8}$ & 546  \\
	10 $ab^{-1}$ (10 TeV) & 4.7543 $\times 10^{9}$ & 21492  \\
	20 $ab^{-1}$ (14 TeV) & 19.0947 $\times 10^{9}$ & 452815\\
		\hline
		 \end{tabular}
    \end{center}
   \end{table}     

    \begin{table}[!htb]
 	  	  \caption{\label{tab4}  Forward-backward asymmetry $A_{FB}$ in $\mu^{+}\mu^{-} \rightarrow W^{+}W^{-} \rightarrow l^{-}\nu l^{+}\overline{\nu}$ collision in case of the different polarization coefficients of $\mu^{-}, \mu^{+}$ beams at 10 TeV. The parameters are chosen as $\Delta k_{\gamma} = -0.135$, $\lambda_{\gamma} = 0.061$, $\Delta k_{Z} = 0.093$, $\lambda_{Z} = -0.062$, $f^{\gamma}_{4} = -0.0012$, $f^{Z}_{4} = 0.001$, $f^{\gamma}_{5} = -0.0012$, $f^{Z}_{5} = 0.001$.} 
 	  	  \vspace*{0.5cm}
 	  	\begin{center}
 	   	  \begin{tabular}{|c |c |c |c| c| c|}     	 
 	   	  \hline
 	   	 $ (P_{\mu^{-}}, P_{\mu^{+}})$ & $A_{FB}$ &  $ (P_{\mu^{-}}, P_{\mu^{+}})$ & $A_{FB}$ &$ (P_{\mu^{-}}, P_{\mu^{+}})$ & $A_{FB}$   \\
 	   	   	\hline  	   
	(-1, -1)& 0.007201 & (0, -1) & 0.00717429 & (1, -1)& 0.005076 \\
		(-1, -0.8) & 0.007198 & (0, -0.8) & 0.00717423 & (1, -0.8) & 0.007018\\
		(-1, -0.6) & 0.007195 & (0, -0.6) & 0.00717422 & (1, -0.6) & 0.007118\\
		(-1, -0.4) & 0.007192 & (0, -0.4) & 0.00719566 & (1, -0.4) & 0.007152 \\
		 (-1, -0.2) & 0.007187 & (0, -0.2) & 0.00719568 & (1, -0.2) & 0.007169 \\
	 	 (-1, 0) & 0.007180 & (0, 0) & 0.00717424 & (1, 0) & 0.007180\\	
	 	 (-1, 0.2) & 0.007169 & (0, 0.2) &  0.00719568 & (1, 0.2) & 0.007187\\
	 	 (-1, 0.4) & 0.007152 & (0, 0.4) & 0.00719566 & (1, 0.4) & 0.007192\\
	 	 (-1, 0.6) & 0.007118 & (0, 0.6) &  0.00717422 & (1, 0.6) & 0.007195\\
	 	 (-1, 0.8) &  0.007018 & (0, 0.8) &  0.00717423 & (1, 0.8) &  0.007198\\
	 	 (-1, 1) & 0.005076 & (0, 1) & 0.00717429 & (1, 1) & 0.007201 \\
			\hline
							 \end{tabular}
    \end{center}
   \end{table}

 \begin{table}[!htb]
 	  	  \caption{\label{tab5}  Forward-backward asymmetry $A_{FB}$ in $\mu^{+}\mu^{-} \rightarrow ZZ \rightarrow l^{-}l^{+}\nu\overline{\nu}$ collision in case of the different polarization coefficients of $\mu^{-}, \mu^{+}$ beams at 10 TeV. The parameters are chosen as $\Delta k_{\gamma} = -0.135$, $\lambda_{\gamma} = 0.061$, $\Delta k_{Z} = 0.093$, $\lambda_{Z} = -0.062$, $f^{\gamma}_{4} = -0.0012$, $f^{Z}_{4} = 0.001$, $f^{\gamma}_{5} = -0.0012$, $f^{Z}_{5} = 0.001$.} 
 	  	  \vspace*{0.5cm}
 	  	\begin{center}
 	   	  \begin{tabular}{|c |c |c |c| c| c|}     	 
 	   	  \hline
 	   	 $ (P_{\mu^{-}}, P_{\mu^{+}})$ & $A_{FB}$ &  $ (P_{\mu^{-}}, P_{\mu^{+}})$ & $A_{FB}$ &$ (P_{\mu^{-}}, P_{\mu^{+}})$ & $A_{FB}$   \\
 	   	   \hline  	   
	(-1, -1)& 0.005434 & (0, -1) & 0.00542703 & (1, -1)& 0.005076 \\
	(-1, -0.8) &0.005433 & (0, -0.8) &  0.00542706 & (1, -0.8) & 0.005376\\
		(-1, -0.6) & 0.005432 & (0, -0.6) & 0.00542707 & (1, -0.6) & 0.005406\\
		(-1, -0.4) & 0.005431 & (0, -0.4) & 0.00542709 & (1, -0.4) & 0.005418 \\
		 (-1, -0.2) & 0.005429 & (0, -0.2) & 0.0054271 & (1, -0.2) & 0.005423 \\
	 	 (-1, 0) & 0.005427 & (0, 0) & 0.0054271 & (1, 0) & 0.005427\\	
		 (-1, 0.2) & 0.005423 & (0, 0.2) & 0.0054271 & (1, 0.2) & 0.005429\\
	 	 (-1, 0.4) & 0.005418 & (0, 0.4) & 0.00542709 & (1, 0.4) & 0.005431\\
	 	 (-1, 0.6) & 0.005406 & (0, 0.6) &  0.00542708 & (1, 0.6) & 0.005432\\
	 	 (-1, 0.8) &  0.005376 & (0, 0.8) &  0.00542706 & (1, 0.8) & 0.005433\\
	 	 (-1, 1) & 0.005076 & (0, 1) & 0.00542703 & (1, 1) & 0.005434 \\
			\hline
							 \end{tabular}
    \end{center}
   \end{table}   
       \hspace*{0.5cm} viii) The forward-backward asymmetry is defined as the asymmetry between the forward and backward cross-sections, which is a potent tool for demonstrating heightened sensitivity to new particles and interactions \cite{jung, acc, bou, pla, naris}. The $A_{FB}$ values in case of the different polarization coefficients of $\mu^{-}, \mu^{+}$ beams at 10 TeV are given in detail in Table.\ref{tab4} for the $\mu^{+}\mu^{-} \rightarrow W^{+}W^{-} \rightarrow l^{-}l^{+} \nu \overline{\nu}$ processes and Table.\ref{tab5} for the $\mu^{+}\mu^{-} \rightarrow ZZ \rightarrow l^{-}l^{+} \nu \overline{\nu}$ processes, respectively. The result shows that the asymmetry is an observable sensitively to the polarization. The asymmetry can reach 0.007201 and 0.005434 with the polarization of $(P_{\mu^{-}}, P_{\mu^{-}}) = (-1, -1); (1, 1)$ for $\mu^{+}\mu^{-} \rightarrow W^{+}W^{-}/ZZ \rightarrow l^{-}l^{+} \nu \overline{\nu}$ processes, respectively. That means the probability of the $l^{-}l^{+} \nu \overline{\nu}$ production moving in the forward direction. It is noteworthy that the calculated asymmetry within the SM framework, under the condition of polarized beams $(P_{\mu^{-}}, P_{\mu^{-}}) = (1, 1)$, yields the value 0.000236 and 0.000233 in cases of $W^{+}W^{-} \rightarrow l^{-}l^{+} \nu \overline{\nu}$ and $ZZ \rightarrow l^{-}l^{+} \nu \overline{\nu}$, respectively. So the asymmetry $A_{FB}$ within the RS model framework including unparticle and anomalous couplings is much larger than that within the SM framework.

\begin{table}[!htb]
 	  \caption{\label{tab6} The cross-sections for the s-, u-, and t- channels are calculated for various collision energy values. The parameters are chosen as $P_{\mu^{-}} = 0.8, P_{\mu^{+}} = -0.8$, $\Delta k_{\gamma} = -0.135$, $\lambda_{\gamma} = 0.061$, $\Delta k_{Z} = 0.093$, $\lambda_{Z} = -0.062$, $f^{\gamma}_{4} = -0.0012$, $f^{Z}_{4} = 0.001$, $f^{\gamma}_{5} = -0.0012$, $f^{Z}_{5} = 0.001$.} 
 	  	  \vspace*{0.5cm}
 	  	\begin{center}
 	  	  \begin{tabular}{|c|c|c|c|c|}     	 
 	 	  \hline
 	 	  Processes & $\sqrt{s}$ (TeV) & $\sigma_{s}$ (fb) & $\sigma_{u}$ (fb) & $\sigma_{t}$ (fb)\\
 	 	  \hline
 	   $\mu^{+}\mu^{-} \rightarrow W^{+}W^{-} \rightarrow l^{-}\nu l^{+}\overline{\nu} $ & 3 & 2691.6 & - & 24162.9 \\
 	   & 6 & 13249.9 & - & 95750.2 \\
 	   & 10 & 50719.9 & - & 265440  \\
 	   & 14 & 137909 & - & 519975  \\
	\hline
	$\mu^{+}\mu^{-} \rightarrow ZZ \rightarrow l^{-}l^{+}\nu\overline{\nu}$ & 3 & 0.00771 & 0.0000967 & 0.0000967  \\
	& 6 & 0.13426 & 0.000382 & 0.000382 \\
	& 10 & 2.14223 & 0.001059 & 0.001059  \\
	& 14 & 22.6251 & 0.002073 & 0.002073 \\
		\hline
		 \end{tabular}
    \end{center}
   \end{table} 
  
\hspace*{0.5cm} ix) Finally, we compare the relative contributions of the s-, u-, t- scattering channels in Table.\ref{tab6}. The cross-section in t-channel within $W^{+}W^{-} \rightarrow l^{-}l^{+}\nu \overline{\nu}$ is larger than that in s-channel, while the cross-sections contributed by the s-channel in $ZZ \rightarrow l^{-}l^{+}\nu \overline{\nu}$ is larger than that in u-, t- channels, respectively.\\
\hspace*{0.5cm} We note that while the present analysis focuses on the $l^{+}l^{-}\nu\bar{\nu}$ channel, alternative decay modes, such as fully hadronic final states, present significant opportunities due to their larger branching fractions \cite{navas}. Furthermore, within the high energy muon collider environment, rare decay signatures, (e.g., into two massless vector particles, three gauge bosons or a gauge boson plus a meson,...) can serve as complementary and potentially probes of new physics. Any observed deviation from SM predictions in these branching ratios would constitute a powerful and high fidelity signal of physics beyond the SM. Consequently, we identify the detailed investigation of these alternative final states as a primary direction for our future researches, aiming to provide a more comprehensive phenomenological assessment.

  \section{Conclusion}
 \hspace*{1cm} In this paper, by using Feynman diagram techniques we have evaluated the influence of the scalar unparticle and polarization on  the $l^{-}l^{+}\nu \overline{\nu}$ production in the final state  through the exclusive decay of ZZ/WW gauge bosons in detail at the multi-TeV  muon colliders in the RS model. The result shows that with fixed collision energies, the cross-sections for $l^{-}l^{+}\nu \overline{\nu}$ production in the final state depend strongly on the parameters of the unparticle physics, muon polarization coefficients, parameters on anomalous couplings and also KK-graviton propagators. With the benchmark signal point $(\Lambda_{U}, d_{U})$ $= (1 $TeV$, 1.9)$, the total cross-sections achieve the maximum value when both of muon beams polarize left or right. In case of the different polarization, the cross section increases as the collision energy increases. The numerical evaluation shows that the cross-section for $l^{-}l^{+}\nu \overline{\nu}$ final states  through the exclusive decay of WW charged bosons is much larger than that of ZZ neutral bosons under the same conditions (about $10^{6}$ times). It is worth noting that with the contribution of new physics in the RS model, the effect is greatly enhanced and the cross-sections for the production of $l^{-}l^{+}\nu \overline{\nu}$ final states can be measured in the future muon collisions. \\
 \hspace*{1cm} The values of the forward-backward asymmetry $A_{FB}$ in case of the different polarization coefficients of $\mu^{-}, \mu^{+}$ beams at 10 TeV are also given in detail. The result shows that the asymmetry is an observable sensitively to the polarization. Morever, the asymmetry $A_{FB}$ within the RS model framework including unparticle and anomalous couplings is much larger than that within the SM framework.
 \\ 
 \hspace*{1cm} Finally, we note that in this work we have only considered on a theoretical basis, other problems concerning experiments for future muon colliders, the reader can see in detail in Ref.\cite{wulzer}, which indicate that with  center-of-mass energy up to tens of TeV, it can provide an unprecedented potential in probing new physics beyond the SM. Our future investigations will be extended to encompass hadronic and rare modes within the muon scatterings, providing a more comprehensive exploration of the model’s experimental signatures.
  \\
{\bf Acknowledgements}: The work is supported in part by the National Foundation for Science and Technology Development (NAFOSTED) of Vietnam under Grant No. 103.01-2023.50.\\
\\
\\
\\

\newpage

  \begin{center} 
 { APPENDIX A: Feynman diagrams for the considered processes}\\
 \end{center} 
\begin{figure} [!htbp]
\begin{center}
\includegraphics[width= 18 cm,height= 6 cm]{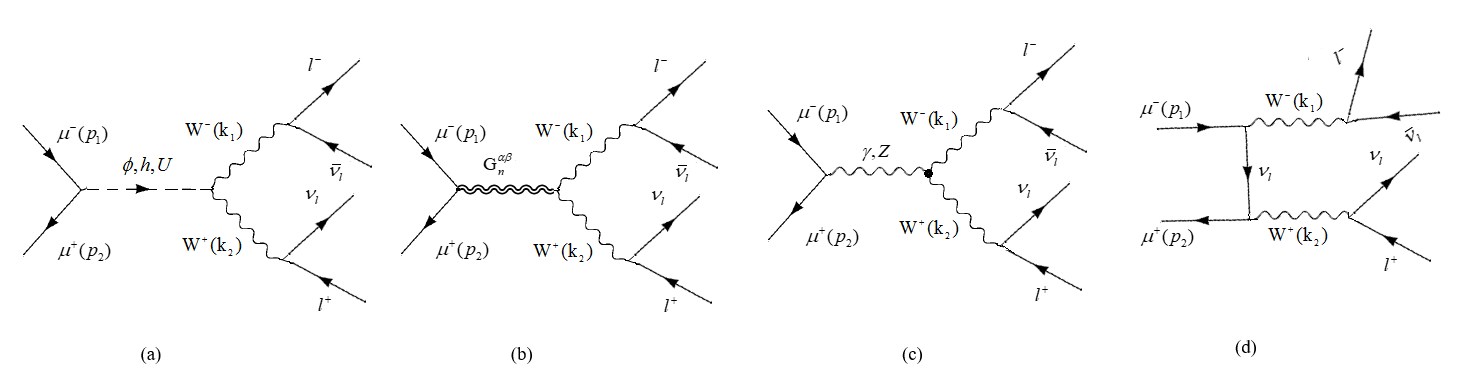}
\caption{\label{Fig.6} Feynman diagrams for $\mu^{+}\mu^{-} \rightarrow  W^{+}W^{-} \rightarrow l^{-}l^{+}\nu \overline{\nu}$ collision.}
\end{center}
\end{figure}
\newpage

\begin{figure} [!htbp]
\begin{center}
\includegraphics[width= 18 cm,height= 10 cm]{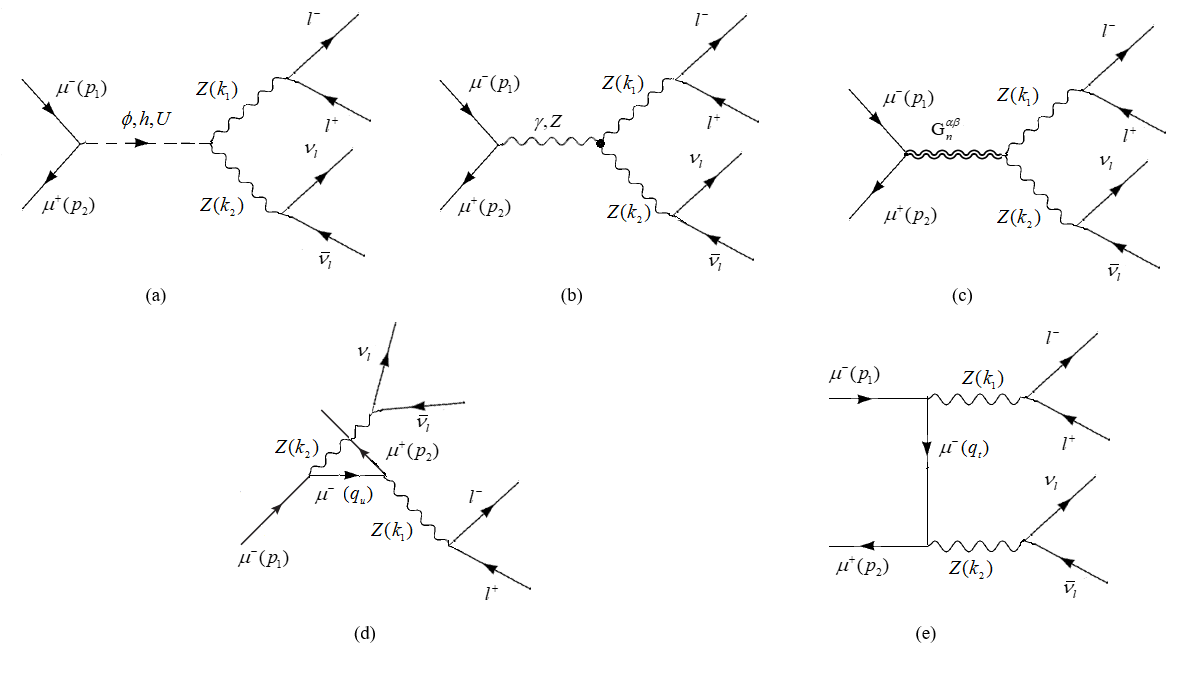}
\caption{\label{Fig.7} Feynman diagrams for $\mu^{+}\mu^{-} \rightarrow  ZZ \rightarrow l^{-}l^{+}\nu \overline{\nu}$ collision.}
\end{center}
\end{figure}
 
  \newpage
  \begin{center} 
 { APPENDIX B: The relevant couplings in the RS model}\\
 \end{center}
\hspace*{1cm} Feynman rules for the vertices in the RS model related to our calculations are listed as below\\
$\overline{\psi} (k_{1}) - \psi (k_{2}) - h (q)$ vertex:
\begin{equation}
i\overline{g}_{f\overline{f}h} = -i\dfrac{gm_{f}}{2m_{W}}\left( d + \gamma b\right),
\end{equation}
$\overline{\psi} (k_{1}) - \psi (k_{2}) - \phi (q)$ vertex:
\begin{equation}
i\overline{g}_{f\overline{f}\phi} = -i\dfrac{gm_{f}}{2m_{W}}\left( c + \gamma a\right),
\end{equation}
$W^{\mu} (k_{1})  - W^{\nu} (k_{2}) - h (q)$ vertex:
\begin{equation}
\begin{aligned}
&i\overline{g}_{Wh}\left[\eta^{\mu\nu} - 2g^{W}_{h}\left(\left(k_{1}k_{2}\right)\eta^{\mu\nu} - k_{1}^{\nu}k_{2}^{\mu}\right)\right]\\
=& igm_{W}\left(d + \gamma b - \gamma b \kappa_{W}\right)\left[\eta^{\mu\nu} - 2g^{W}_{h}\left(\left(k_{1}k_{2}\right)\eta^{\mu\nu} - k_{1}^{\nu}k_{2}^{\mu}\right)\right],
\end{aligned}
\end{equation}
$W^{\mu} (k_{1})  - W^{\nu} (k_{2}) - \phi (q)$ vertex:
\begin{equation}
\begin{aligned}
&i\overline{g}_{W\phi}\left[\eta^{\mu\nu} - 2g^{W}_{\phi}\left(\left(k_{1}k_{2}\right)\eta^{\mu\nu} - k_{1}^{\nu}k_{2}^{\mu}\right)\right]\\
= &igm_{W}\left(c + \gamma a - \gamma a \kappa_{W}\right)\left[\eta^{\mu\nu} - 2g^{W}_{\phi}\left(\left(k_{1}k_{2}\right)\eta^{\mu\nu} - k_{1}^{\nu}k_{2}^{\mu}\right)\right],
\end{aligned}
\end{equation}
$Z^{\mu} (k_{1}) - Z^{\nu} (k_{2}) - h (q)$ vertex:
\begin{equation}
i\overline{g}_{hZ}\left[\eta^{\mu\nu} - 2g^{Z}_{h}\left(\left(k_{1}k_{2}\right)\eta^{\mu\nu} - k_{1}^{\nu}k_{2}^{\mu}\right)\right],
\end{equation}
$Z^{\mu} (k_{1}) - Z^{\nu} (k_{2}) - \phi (q)$ vertex:
\begin{equation}
i\overline{g}_{\phi Z}\left[\eta^{\mu\nu} - 2g^{Z}_{\phi}\left(\left(k_{1}k_{2}\right)\eta^{\mu\nu} - k_{1}^{\nu}k_{2}^{\mu}\right)\right],
\end{equation}
$\overline{\psi} (k_{1}) - \psi (k_{2})  - U (q)$ vertex:
\begin{equation}
i\overline{g}_{f\overline{f}U} = i\dfrac{\lambda_{ff}}{\Lambda_{U}^{d_{U} - 1}},
\end{equation}
$\gamma^{\mu} (k_{1}) - \gamma^{\nu} (k_{2}) - U(q)$ vertex:
\begin{equation}
-i\overline{g}_{\gamma\gamma U}\left[(k_{1}k_{2})\eta^{\mu\nu} - k_{1}^{\nu}k_{2}^{\mu}\right] = -4i\dfrac{\lambda_{\gamma\gamma}}{\Lambda_{U}^{d_{U}}} \left[(k_{1}k_{2})\eta^{\mu\nu} - k_{1}^{\nu}k_{2}^{\mu}\right],
\end{equation}
$W^{\mu} (k_{1}) - W^{\nu} (k_{2}) - U(q)$ vertex:
\begin{equation}
-i\overline{g}_{WW U}\left[(k_{1}k_{2})\eta^{\mu\nu} - k_{1}^{\nu}k_{2}^{\mu}\right] = -4i\dfrac{\lambda_{WW}}{\Lambda_{U}^{d_{U}}} \left[(k_{1}k_{2})\eta^{\mu\nu} - k_{1}^{\nu}k_{2}^{\mu}\right],
\end{equation}
$Z^{\mu} (k_{1}) - Z^{\nu} (k_{2}) - U(q)$ vertex:
\begin{equation}
-i\overline{g}_{ZZU}\left[(k_{1}k_{2})\eta^{\mu\nu} - k_{1}^{\nu}k_{2}^{\mu}\right] = -4i\dfrac{\lambda_{ZZ}}{\Lambda_{U}^{d_{U}}}\left[(k_{1}k_{2})\eta^{\mu\nu} - k_{1}^{\nu}k_{2}^{\mu}\right]. 
\end{equation}
Here $\gamma = \upsilon /\Lambda, \upsilon = 246$ GeV, $a = -\dfrac{cos\theta}{Z}, b = \dfrac{sin\theta}{Z}, c = sin\theta + \dfrac{6\xi\gamma}{Z}cos\theta, d = cos\theta - \dfrac{6\xi\gamma}{Z}sin\theta$, $\theta$ is the mixing angle, $g^{W}_{h} = \dfrac{\gamma b}{(d + \gamma b - \kappa_{W}\gamma b)m^{2}_{W}}\left(\dfrac{1}{2kb_{0}} + \dfrac{\alpha b_{2}}{8\pi sin^{2}\theta_{W}}\right)$, $\kappa_{W} = \dfrac{3m^{2}_{W}kb_{0}}{2\Lambda^{2}(k/M_{Pl})^{2}}$, $\dfrac{1}{2}kb_{0} \sim 35$, $g^{W}_{\phi} = \dfrac{\gamma a}{(c + \gamma a - \kappa_{W}\gamma a)m^{2}_{W}}\left(\dfrac{1}{2kb_{0}} + \dfrac{\alpha b_{2}}{8\pi sin^{2}\theta_{W}}\right)$ \cite{ahm}, $b_{3} = 7, b_{2}= 19/6, b_{Y} = - 41/6$, $\theta_{W}$ stands for the Weinberg angle. \\
\hspace*{1cm} The Feynman rule corresponding to the interaction of two SM Dirac fermions of mass $m_{\psi}$ with one KK-graviton $G_{n}^{\mu\nu}(q) - \overline{\psi} (k_{1}) - \psi (k_{2})$ is given by
\begin{equation}
-\frac{i}{4\Lambda} \left[\gamma^{\mu} (k^{2\nu} - k^{1\nu}) + \gamma^{\nu} (k^{2\mu} - k^{1\mu}) - 2 \eta^{\mu\nu} (\widehat{k}_{2} - \widehat{k}_{1} - 2m_{\psi} )\right].
\end{equation}
\hspace*{1cm} The interaction between two SM gauge bosons of mass $m_{A}$ and one KK-graviton $G_{n}^{\mu\nu}(q) - A^{\alpha} (k_{1}) - A^{\beta} (k_{2})$ is given by
\begin{equation}
-\frac{i}{\Lambda} \left(m^{2}_{A} C^{\mu\nu\alpha\beta} + W^{\mu\nu\alpha\beta}\right).
\end{equation}
\hspace*{1cm} The propagator of the n-th KK-graviton mode, with mass $m_{G_{n}}$, decay width $\Gamma_{n}$ and 4-momentum $q$ in the unitary gauge is:
\begin{equation}
i\Delta^{G}_{\mu\nu\alpha\beta} = \dfrac{iP_{\mu\nu\alpha\beta} (q, m_{G_{n}})}{q^{2} - m_{G_{n}}^{2} + i m_{G_{n}} \Gamma_{n}}.
\end{equation}

\end{document}